\documentclass[12pt]{article} 
\title{Ultra-High Energy Cosmic Rays from Galactic Supernovae}  
\author{{\it Richard Shurtleff~}\thanks{affiliation and mailing 
address: Department of Science, 
Wentworth Institute of Technology, 550 Huntington Avenue, 
Boston, MA, USA, ZIP 02115, telephone number: (617) 989-4338, e-mail address: shurtleffr@wit.edu}} 
\begin{document} 
          
\maketitle 

\begin{abstract} 

Suppose that even the highest energy cosmic rays (CRs) observed on Earth are protons accelerated in local Milky Way Galaxy sources, with few if any from more distant sources. In this paper we treat the problem that supernovae remnants likely produce protons with energies up to about a PeV, but CRs with 100s of EeV energy are observed. We assume with minimal comment the idea that `new physics' is at work and we accept that a CR's collision energy at the Earth exceeds its kinetic energy as it travels through the Galaxy. There is some evidence that the collision energy-kinetic energy difference has been seen at the Tevatron and LHC, but it is small enough to attribute to standard physics. This sets the threshold for energy bifurcation. Based on this threshold and the CR spectrum endpoint, a formula for collision energy as a function of kinetic energy is derived. With the function and the observed CR spectrum we can predict the average spectrum of CR sources. Also we can estimate the collision energies of proton beams as terrestrial particle accelerators advance and produce beams with higher kinetic energies.

\vspace{0.5cm}
 
\vspace{0.5cm}
%96.50.S- 	Cosmic rays ; 	98.70.Sa 	Cosmic rays (including sources, origin, acceleration, and interactions) ; 98.38.Mz 	Supernova remnants ; 14.20.Dh 	Protons and neutrons 

\end{abstract}
\pagebreak
\section{Introduction} \label{into}

If we assume a local origin for cosmic rays, then a cosmic ray (CR) is accelerated in the Galaxy and has a trajectory through the Galaxy. 

We consider CR protons that are accelerated in supernovae remnants (SNRs). Acceleration of charges in SNR shockwaves and trajectories in the Galaxy are well researched and continue to be subjects of ongoing investigations. The physics of both acceleration and trajectory accommodate a kinetic energy of at most 1 PeV or so.\cite{PeV1,PeV2}

The most energetic CR protons deposit hundreds of EeVs into the Earth's atmosphere, so-called ultrahigh energy cosmic rays (UHECRs). We assume that collisions between a CR proton and an atmospheric nucleus involve the same forces that have been studied at lower energies at terrestrial particle accelerators. Thus the UHECRs deposit 100s of EeV, yet are accelerated to at most 1 PeV in SNR shockwaves. With conventional forces for a CR's collision with an atmospheric nucleus, with conventional forces for its acceleration at an SNR, and with conventional forces for its trajectory in the Galaxy, one remaining possibility is to contemplate the conjecture that a CR proton with a given kinetic energy carries also a collision energy that exceeds its kinetic energy.

Protons with 1 PeV kinetic energy are beyond the reach of controlled experiments, so we are free to invoke `new physics'; a new relationship is presumed to exist between the collision energy and kinetic energy of protons, both measurable quantities. The two energies are measured differently and so may themselves differ. Kinetic energy depends on mass, timing and displacement, whereas collision energy depends on summing the individual energies of the collision products. The collision energy cannot be measured before the collision and the kinetic energy cannot be measured after.

There are ways to make sense of collision energy exceeding kinetic energy. One way uses excited states. In a collision any incident particle in an excited state can deliver its kinetic energy and some or all of its excited state energy. The CR situation suggests that a proton has more internal energy the faster it goes. And that suggests a special frame in which the proton at rest has zero excess internal energy; the special frame is perhaps the cosmic microwave background (CMB) frame. So the `new physics' could explain the extra energy, the collision energy minus the kinetic energy, and explain how that energy depends on speed in the CMB frame. Now turn from speculation to experimental results.

Terrestrial particle accelerators produce proton beams with kinetic energies of about 1 TeV or so, with 7 TeV the design maximum for the Large Hadron Collider (LHC). From the LHC and the earlier Tevatron, there have been reports of excessive secondary particle production and unexpectedly high charged particle multiplicities.\cite{ExtraMesons}-\cite{Aad2011} These unexpected results could be interpreted as showing proton collision energies exceed kinetic energies. Instead, these small effects are mainly interpreted as modifying and improving accepted theories. 

Here we disagree and interpret the results as new physics.  We take a convenient percentage, 10\% at 1 TeV, as the fractional difference of collision and kinetic energy. This sets the threshold for the onset of collision/kinetic energy bifurcation.

In Sec. 2 we discuss some of the available data at the CR spectrum endpoint and at the LHC-Tevatron threshold. In Sec. 3 we assume the fractional energy difference is proportional to some power of the kinetic energy. By applying the data from Sec. 2, we deduce a formula for collision energy as a function of kinetic energy. In Sec. 4 we discuss the formula and suggest a way to test it.

\section{Data} \label{data}

Consider the acceleration of protons in individual supernova remnants (SNRs). Some of these protons collide with ambient matter with sufficient energy to produce neutral pions that then decay producing high energy gamma rays. Since gammas are not charged, their incoming direction when detected points back to the source. For example, the spectrum of very high energy gamma rays from the SNR RX J1713.7-3946 has been measured.\cite{HESS,CANGAROO-II} The end-of-spectrum for the gammas occurs at about 40 TeV. Applying the rule that the primary protons have energies of about 10 times the observed gammas,\cite{rule} we see that the protons have energies of up to about 400 TeV or so. We round that up and take 1 PeV as the maximum energy of a proton accelerated in an SNR. 

Consider the trajectories of protons. One expects that 1 PeV protons travel in tight one parsec radius spirals along the $10^{-10}$~T Galactic magnetic field lines.\cite{galMag} Estimates of diffusion coefficients and escape times are ongoing topics.\cite{Ptuskin} The magnetic field effectively randomizes the motion of 1 PeV protons and the arrival direction at the Earth of a CR proton is not related to the direction of its source SNR. SNR accelerated CR protons arrive nearly isotropically distributed over the sky.

The distinction between kinetic energy $E_{K}$ and total energy is not important for the high energies considered in this paper. In theory, $E_{K}$ can be determined for a specific proton by measuring time-of-flight over a given distance or by measuring the orbit radius in a known magnetic field.  It is not practical to measure $E_{K}$ for CR protons. From observations of gamma rays at SNRs, about 1 PeV is the maximum energy of a CR proton accelerated in an SNR. For most CR protons arriving at the Earth, this initial energy is not likely to change very much along its trajectory, so we have $E_{K} \leq$ 1 PeV.

Kilometer scale observatories search the sky for showers of particles and photons produced when a high energy cosmic ray encounters an atmospheric nucleus. The `collision energy' $E_{C}$ of the CR is then the sum of the energies of these particles, the total energy deposited into the atmosphere by the CR. Observations put the endpoint of the collision energy spectrum at about 100 EeV.\cite{1015,1015a} These CRs are the ultrahigh energy cosmic rays.

The most energetic protons observed are the most energetic protons that have been accelerated in Galactic SNRs. Therefore at the end of the CR spectrum, $E_{C} \approx$ $10^5 E_{K}.$

Protons have been accelerated to a TeV and beyond at the Tevatron and the LHC. In contrast with cosmic rays, at particle accelerators it is practical to measure time and distance so $E_{K}$ is measured. Unseen neutrinos and other undetected particles make it a challenge to determine $E_{C}$ directly. Even though $E_{C}$ is not measured directly there are some indications from the data that collision energies are larger than kinetic energies.

For example one experiment at the Tevatron reports: `This selection yields a sample of 34~367 J/$\psi$ candidates, where the estimated number of real J/$\psi$ mesons is 32~642 $\pm$ 185.'\cite{ExtraMesons} The experiment was running at a kinetic energy of $E_{K}$ = 0.900 TeV. Thus it would seem that $E_{C}$ is about 5\% more than $E_{K}$ at 0.900 TeV. Complicating the situation, however, is the fact that the estimate of 32~642 mesons is based on lower energy runs and if there is a collision energy/kinetic energy effect at the lower energy, then that complicates the interpretation.

The J/$\psi$ example is typical of the data we select from experiments at the Tevatron and LHC. An excess is observed at the start of a higher energy run and it is compared with an estimate based on lower energy runs. But at the lower energy runs models based on the data would have been massaged for accuracy, thereby canceling the effect at the lower energies. Thus we cannot simply translate the excess into a value for the collision minus kinetic energy difference.

Many sources report unexpectedly high charged particle multiplicities.\cite{Ghosh2012}-\cite{Aad2011} The job of estimating charged particle multiplicities is given to various models, each of which contain parameters that are chosen to fit the data. `PYTHIA' and `PHOJET' are two of these models,\cite{PP} sometimes called Monte-Carlo generators. A `tune' is a version with parameters selected to fit experimental outcomes. So, as the proton energy increases, newer and newer tunes are created to describe the swarms of particles created in proton-proton collisions at the Tevatron and now at the LHC. `Unexpectedly high multiplicities' means that the tunes set at lower proton energy predict fewer particles than are actually produced at higher proton energies. 

To summarize the cited reports, consider the following quote from the most recent article: 
\begin{quote}
  ``At LHC, multiplicity distributions in proton-proton
collisions at center-of-mass energies $\sqrt{s}$ = 0.9, 2.36 and
7 TeV have been measured by different experiments, in
different kinematic ranges and for different classes of
events. All these LHC-experiments find that the mean
multiplicities at the new LHC energies ($\sqrt{s}$ = 2.36 and 7
TeV) had been underestimated by the event generators
(like PYTHIA, PHOJET etc.) in use.''\cite{Ghosh2012}
 \end{quote}
The other references, \cite{CDF2011}-\cite{Aad2011}, make similar comments.

Since multiplicity increases with collision energy, reports of unexpectedly high multiplicities could indicate an unexpectedly high collision energy. But the collision energy $E_{C}$ available to a proton is one parameter that is not adjusted when fitting phenomenological models like PYTHIA and PHOJET. The collision energy is assumed to be the same as the kinetic energy which is known because the location of each bunch of protons in the accelerator is known as a function of time. Thus, the unexpected results could be interpreted as showing collision energies exceed kinetic energies. Instead, they are interpreted as modifying accepted theories.

If experiments at the LHC are just beginning to see collision energies increase above kinetic energies, then that determines a threshold, perhaps $E_{C}$ exceeds $E_{K}$ by 10\% at 1 TeV. In the following section, combining the data at the 1 TeV threshold with the data at the ultrahigh energy end of the CR spectrum, we deduce a formula for collision energy $E_{C}$ as a function of kinetic energy $E_{K}.$

\section{The Formula} \label{formula}

Given the data from the last section, in this section we obtain a phenomenological formula for the dependence of the collision energy $E_{C}$ of a proton as a function of its kinetic energy $E_{K}.$ $E_{C}$ is the energy available to an incident proton in a collision, and can be measured by summing the energies of the collision products. 

Suppose, for simplicity, that the dependence of collision energy on kinetic energy takes the form 
\begin{equation} \label{formula1} 
															E_{C} = E_{K}(1 + a \gamma^{n})															\, ,
\end{equation} 
where $a$ and $n$ are real-valued constants. The ratio of kinetic energy to rest energy defines the relativistic $\gamma,$ $\gamma \equiv$ $E_{K}/mc^2$ . With other constants $b$ and $m,$ we could have written (\ref{formula1}) as $E_{C} - E_{K}$ = $b \, E_{K}^{m},$ which is just a simple power law for the difference of $E_{C}$ and $E_{K}.$ 

We seek values for $a$ and $n.$ To determine $a$ and $n$ we need two sets of data. At the end of the CR spectrum, a UHECR is accelerated and travels in the Galaxy with a kinetic energy of about 1 PeV, $E_{K  1}$ = $1 \times 10^{15}$ eV. In the upper atmosphere of the Earth, the UHECR deposits a collision energy of 100 EeV, $E_{C 1}$ = $1 \times 10^{20}$ eV. Thus we have $E_{C}$ and $E_{K}$ at the high energy end of the CR spectrum. 

At the onset of bifurcation, we assume that there is a 10\% effect at a kinetic energy of 1~TeV, so we have $E_{K  0}$ = $1 \times 10^{12}$ eV and $E_{C 0}$ = $1.1 \times 10^{12}$ eV.  

Substituting the two sets of data in (\ref{formula1}) gives
\begin{equation} \label{formula2}
															10^{5} =  a 10^{6n} \quad {\mathrm{and}} \quad 10^{-1} = a 10^{3n} \, ,
\end{equation}
where we drop one compared to $10^5$ and we approximate the rest energy, $mc^2 \approx$ $1 \times 10^9$ eV.  We find that 
\begin{equation} \label{aANDn}
																a = 1 \times 10^{-7} \quad {\mathrm{and}} \quad n = 2 										\, .
\end{equation}
Thus we get a formula,
\begin{equation} \label{formula3} 
															E_{C}  = E_{K}(1 + 1 \times 10^{-7} \gamma^{2}) = E_{K}\left(1 + 1 \times 10^{-7} \, \frac{E_{K}^{2}}{m^2c^4}\right)														\, ,
\end{equation}
for the dependence of collision energy $E_{C}$ on kinetic energy $E_{K}.$ The fractional difference of the two types of energies increases with the square of the energy.

 \section{Discussion} \label{discuss}
 
 The observed CR flux $dN/dE_{C}$ must be understood as depending on the collision energy $E_{C}$ which is measured rather than the kinetic energy which is not measured. Now, given (\ref{formula3}), we can predict the spectrum of CRs as a function of kinetic energy $E_{K},$
\begin{equation} \label{dNdE1} 
															\frac{dN}{dE_{K}}  = \frac{dN}{dE_{C}} \frac{dE_{C}}{dE_{K}} = \frac{dN}{dE_{C}} \left(1 +  3 \times 10^{-7} \, \frac{E_{K}^{2}}{mc^2} \right)			\, .
\end{equation}
Over many decades of energy from $E_{C}$ = $3 \times 10^{12}$ eV up to about $10^{20}$ eV the CR spectrum has a spectral index of about 3, $dN/dE_{C} \propto$ $E_{C}^{-3}.$ (We have rounded off the famous `knee' at $E_{C}$ = $4 \times 10^{15}$ eV where the spectral index changes from 2.7 to 3.1.) Then, by (\ref{formula3}) and (\ref{dNdE1}) for high energy CRs, $E_{C} >$  $3 \times 10^{12},$ the spectral index for $dN/dE_{K}$ is about 7,
\begin{equation} \label{dNdE2} 
					\frac{dN}{dE_{C}}  \propto E_{C}^{-3} \quad {\mathrm{and}} \quad \frac{dN}{dE_{K}}  \propto E_{K}^{-7} 			\, . 
\end{equation} 
Since a CR proton leaving a supernova remnant has a kinetic energy $E_{K},$ the $dN/dE_{K}$ in (\ref{dNdE2}) is the spectrum for the acceleration of protons averaged over all sources. The steepness indicated by a spectral index of 7 may say something about SNR sources. The fraction of all SNR sources that contribute CRs at a given energy seems to decrease rapidly with energy for the highest energy CRs.

 Particle states and particle fields differ. States transform differently than fields. Conventionally, under a general combination of a rotation and boost followed by a translation of spacetime, particle states transform via a unitary rep, while  particle fields transform nonunitarily.\cite{W}  Collisions change the state of a particle, so perhaps states determine collision energies. And if the phase of a particle field is extreme along the most-likely trajectory, then perhaps fields determine trajectories and kinetic energies. So, if states and fields differ in new ways, then among the consequences may be new relationships between collision energy and kinetic energy. Such a point of view is embodied in a calculation.\cite{SDualPhase} Among the results of the calculation is formula (\ref{formula1}) with $n$ = 2 and $a $ = $4.2 \times 10^{-8},$ which agrees nicely with (\ref{aANDn}). I mention this to show that it is not unthinkable that collision and kinetic energies differ.
 
We also predict that pp-collisions for the maximum design energy at the LHC should show clearly that collision energy exceeds kinetic energy. The LHC is designed to operate at a kinetic energy of $E_{K}$ = 7 TeV. By (\ref{formula3}), the collision energy for such a proton is $E_{C}$ = 40 TeV. With $a$ = $4.2 \times 10^{-8},$ we get $E_{C}$ = 20 TeV. Both 20 TeV and 40 TeV are much greater than 7 TeV. Even if $a$ is actually somewhat smaller, the LHC at maximum design energy should produce collisions with enough energy so that adding up the energies of collected particles and estimating the energies of the particles not collected, i.e. measuring the collision energy as directly as possible, would give a result that exceeds the kinetic energy, $E_{C} >$ $E_{K}.$ If the test can be performed accurately, then it should illuminate the relationship between a proton's collision energy and its kinetic energy.

 \end{document}